\documentclass[tradiabstract]{aa}
\usepackage{graphicx}
\usepackage{txfonts}

\begin{document}
\title{BD+43$\degr$ 3654 -- a blue straggler?}

\author{V.V.~Gvaramadze\inst{1,2}
\and D.J.~Bomans\inst{1}}

\institute{
Astronomical Institute, Ruhr-University Bochum, Universit\"{a}tstr. 150, 
44780 Bochum, Germany
\thanks{\email{vgvaram@sai.msu.ru; bomans@astro.rub.de}}
\and
Sternberg Astronomical Institute, Moscow State University, 
Universitetskij Pr. 13, Moscow 119992, Russia\\
}

\date{Received 2 April 2008 / Accepted 22 May 2008}

\abstract{The astrometric data on the runaway star BD+$43\degr \,
3654$ are consistent with the origin of this O4If star in the center
of the Cyg OB2 association, while BD+$43\degr \, 3654$ 
is younger than the association. To reconcile this discrepancy, 
we suggest that BD+$43\degr \, 3654$ is a blue straggler formed via 
a close encounter between two tight massive binaries in the core of
Cyg\,OB2. A possible implication of this suggestion is that the very massive (and therefore apparently very young) stars in Cyg\,OB2 could be
blue stragglers as well. We also suggest that the binary-binary encounter producing BD+$43\degr \, 3654$ might be responsible for ejection of two
high-velocity stars (the stripped helium cores of massive stars) -- the 
progenitors of the pulsars B2020+28 and B2021+51.}


 \maketitle

\titlerunning{BD+43$\degr$ 3654 -- a blue straggler?}
\authorrunning{Gvaramadze \& Bomans}

\section{Introduction}
%
\object{BD+$43\degr \, 3654$} is a massive (O4If) runaway star
located $\sim 2{\fdg}7$ northeast of the \object{Cyg OB2}
association (Comer\'{o}n \& Pasquali \cite{co07}). The astrometric data on
BD+$43\degr \, 3654$ suggest that this star was ejected from the
center of Cyg\,OB2 $\sim 1.8$ Myr ago (Comer\'{o}n \& Pasquali \cite{co07}). 
BD+$43\degr \, 3654$ is one of the three known very massive ($\ga
60$ M$_{\odot}$) runaway stars (the two other stars are \object{$\zeta$
Pup} and \object{$\lambda$ Cep}) whose high ($\ga 40 \, {\rm km} \,
{\rm s}^{-1}$) peculiar velocities cannot be explained within
the framework of the binary-supernova scenario (see Sect.\,3). The
more likely channel for production of these high-velocity massive
objects is through dynamical processes in the
dense cores of young massive star clusters (Poveda et al. \cite{po67};
Leonard \& Duncan \cite{le90}), particularly through close encounters between 
tight massive binary stars (Mikkola \cite{mi83}; Leonard \& Duncan
\cite{le90}). The most common outcome of binary-binary
encounters is the exchange of the more massive binary components into
a new eccentric binary or a single merged star and ejection of the
less massive ones with high velocities (e.g. Leonard \cite{le95}). The ejection velocity of the lightest components could be as
high as the escape velocity from the surface of the most massive
star in the binaries (i.e. $\sim$ 1000 ${\rm km} \, {\rm
s}^{-1}$; Leonard \cite{le91}). Correspondingly, the recoil velocity of the
newly formed object (the massive binary or the merged star) could
significantly exceed the escape velocity from the potential well of
the parent cluster, so that this object becomes a runaway.

In this Letter we suggest that BD+$43\degr \, 3654$ is a merged
star formed via a close encounter between two tight massive binaries
in the core of Cyg\,OB2 (cf. Leonard \cite{le95}). We suggest that
BD+$43\degr \, 3654$ is a blue straggler, i.e. a rejuvenated star
with the apparent age smaller than the age of the parent association
(see Sect.\,2 and Sect.\,4). The high linear momentum imparted to
BD+$43\degr \, 3654$ implies that the binary-binary encounter was
very energetic and that the lower mass components of the binaries
involved in the encounter were ejected with high velocities.

Interestingly, Cyg OB2 may be associated with the origin of two
high-velocity pulsars, \object{PSR B2020+28} and \object{PSR B2021+51}. 
The high-precision proper motion and parallax measurements for these 
pulsars (presently separated by $\sim 23\degr$) lead to the suggestion that they originated $\sim 1.9$ Myr ago within several parsecs of 
each other in the direction of Cyg OB2 (Vlemmings et al. \cite{vl04}). 
Vlemmings et al. (\cite{vl04})
believe that the progenitors of both pulsars were members of a
common binary and that the pulsars were separated at the
birth of the second one, following asymmetric supernova
explosion. An alternative possibility is that the pulsars were 
separated before their birth, and that they are the remnants of 
runaway stars ejected from Cyg\,OB2 due to the dynamical three- or four-body encounters (Gvaramadze \cite{gv07}, Paper\,I). The relative position 
of the pulsars and BD+$43\degr \,3654$ on
the sky and the similarity between their kinematic ages suggest that the three objects might be ejected
from Cyg\,OB2 via the same dynamical event -- the close encounter between two massive binaries (Sect.\,4; cf. Gies \& Bolton \cite{gi86}; Hoogerwerf et al. \cite{ho01}; Gualandris et al. \cite{gu04}).

\section{BD+43$\degr$ 3654 and Cyg OB2}

The isochronal age 
of BD+$43\degr \, 3654$, derived by Comer\'{o}n \& Pasquali (\cite{co07}) from the position of this O4If star in the Hertzsprung-Russell (HR) diagram 
and the isochrones from the evolutionary models for massive stars by Meynet et al. (1994), is similar to the kinematic age (i.e. the time 
since the ejection of the star from Cyg\,OB2).
Both ages should be compared with the age of the parent association
of $\sim$ 1-5 Myr (e.g. Bochkarev \& Sitnik \cite{bo85}; Torres-Dodgen et
al. \cite{to91}; Herrero et al. \cite{he99}; Hanson \cite{ha03}; Kn\"{o}dlseder \cite{kn03}).
The wide age spread could be considered as an indication that the
star formation in Cyg\,OB2 is non-coeval (Massey \& Thompson \cite{ma91};
Hanson \cite{ha03}). On the other hand, the youngest ages of $\la 3$ Myr
come from the presence in Cyg\,OB2 of a two dozen early type O
stars, while the evolutionary status of the less massive stars in
the association is consistent with an age of 4-5 Myr (the
situation typical of young massive star clusters and associations;
e.g. Massey \cite{ma03}). One can hypothesize that the actual age of
Cyg\,OB2 is $\sim$ 4-5 Myr and that the most massive (and therefore
apparently the youngest) stars in the association are in fact the
rejuvenated stars (blue stragglers) formed via merging of less
massive stars in the course of close binary-binary encounters $\sim 2$ 
Myr ago, when Cyg\,OB2 was much more compact (see below). 
In our reasoning we proceed from the results of binary-binary
scattering experiments by Leonard (\cite{le95}), which showed that about
half of the merged stars formed via close encounters between binaries
remain bound to the parent cluster (some of them form binaries
with another blue straggler) and appear much younger than other
members of the cluster (see also Portegies Zwart et al. \cite{po99}).

An indirect support for our hypothesis comes from the recent study
of early-A stars in the direction of Cyg\,OB2 by Drew et al. (\cite{dr08}).
This study revealed several hundred A stars
within the boundaries of the association and suggested that the age
of these stars is $\geq 5$ Myr, provided that they are located at a
distance of $\leq 1.7$ kpc (i.e. at the distance of Cyg\,OB2;
Kiminki et al. \cite{ki07}; cf. Torres-Dodgen et al. \cite{to91}; Massey \&
Thompson \cite{ma91}). The age of $\sim$ 3-5 Myr would also be required if:
(i) the pulsars B2020+28 and B2021+51 indeed originate in
Cyg\,OB2 (Paper\,I; see also Sect.\,4), (ii) the HEGRA TeV source
observed in the direction of Cyg\,OB2 (Aharonian et al. \cite{ah02}) is
related to a young pulsar in the association (Bednarek \cite{be03}) and
(iii) a partially non-thermal shell-like object coincident with the
HEGRA source is a supernova remnant (Butt et al. \cite{bu08}).

Moreover, the stellar content of Cyg\,OB2 could be contaminated by
young massive stars injected into the association from nearby
associations (Uyaniker et al. \cite{uy01}) and the numerous young clusters
around Cyg\,OB2\footnote{A study of the \object{Cygnus\,X}
region by Schneider et al. (\cite{sc06}) suggests that these clusters and
the Cyg\,OB2 and OB9 associations form a coherent complex and that
formation of the clusters was triggered by the effect of massive
stars in the associations (cf. Le Duigou \& Kn\"{o}dlseder \cite{le02}).}
(Dutra \& Bica \cite{du01}; Comer\'{o}n \& Torra \cite{co01}; Le Duigou \&
Kn\"{o}dlseder \cite{le02}). In the latter case, some of the `alien' O
stars can be produced through merging of two or three stars in the
course of binary-binary encounters in clusters of B-type stars (see
Leonard \cite{le95}). For example, in our search
for bow shocks around Cyg\,OB2 (Gvaramadze \&
Bomans, in preparation), we discovered a bow shock produced by one
of the early-type stars from the list of new members of the
association by
Comer\'{o}n et al. (\cite{co02}). This star [designated by Comer\'{o}n et
al. (\cite{co02}) as A37] was classified by Hanson (\cite{ha03}) as a O5V star,
and therefore it should be a young ($\la 1$ Myr) object. The
photometric distance to this star of 1.7-1.8 kpc [derived with use
of the UBVJHK synthetic photometry of Galactic stars by Martins \&
Plez (\cite{ma06})] is consistent with the distance to Cyg\,OB2. The
astrometric data on the star and the geometry of the bow shock,
however, suggest that this high-velocity ($\sim$ 120 ${\rm km} \,
{\rm s}^{-1}$) runaway was ejected either from the young cluster
embedded in the H\,{\sc II} region \object{DR\,15} (located on the
border of Cyg\,OB2 $\sim 1\degr$ west of the current position of
the star) or from the open cluster \object{NGC\,6913}
(centered $\sim 3{\fdg}4$ west of the star).

At the distance to Cyg\,OB2 of $\sim 1.7$ kpc,
the peculiar (transverse) velocity of BD+$43\degr \, 3654$ is 
$\simeq 40\pm 10 \,
{\rm km} \, {\rm s}^{-1}$ [we used here the Galactic constants 
$R_0 = 8$ kpc and $\Theta _0 =200 \, {\rm km} \, {\rm s}^{-1}$ 
(e.g. Reid \cite{re93}; Kalirai et al. \cite{ka04}; Avedisova \cite{av05}) and the solar
peculiar motion $(U_{\odot} ,V_{\odot} , W_{\odot} )=(10.00, 5.25, 7.17)
\, {\rm km} \, {\rm s}^{-1}$ (Dehnen \& Binney \cite{de98}); cf. Comer\'{o}n \& Pasquali (\cite{co07})]. 
The position of BD+$43\degr \, 3654$ in the HR diagram and the 
evolutionary tracks by Meynet et al. (\cite{me94}) imply an initial mass 
of the star of $\simeq 70\pm 15 \, M_{\odot}$ (Comer\'{o}n \& Pasquali
\cite{co07}). The high linear momentum attained by 
this runaway star could be used to constrain the possible mechanisms
of its origin. In Sect.\,3 we show that runaways of this mass and
velocity are unlikely to be produced via the disruption of a binary
due to the (asymmetric) supernova explosion (the binary-supernova
scenario; Blaauw \cite{bl61}; Stone \cite{st91}). Another possibility is that this
star attained a high peculiar velocity via the strong dynamical
three- or four-body encounter (the dynamical ejection scenario;
Poveda et al. \cite{po67}; Gies \& Bolton \cite{gi86}). In Sect.\,4, we suggest
that the most likely path for the origin of BD+$43\degr \, 3654$
is through the close encounter between two tight massive binaries.

Cyg\,OB2 is one of the most compact and massive associations in the
Milky Way. It contains $\sim$ 100 O stars or stars with O-type
progenitors (Kn\"{o}dlseder \cite{kn00}; Comer\'{o}n et al. \cite{co02}). The half
light radius of Cyg\,OB2 is $\sim 6$ pc (Kn\"{o}dlseder \cite{kn00}). Assuming that the association expands with a
velocity equal to its velocity dispersion ($\sim 2.4 \, {\rm km} \,
{\rm s}^{-1}$; Kiminki et al. \cite{ki07}), one finds that the majority of
massive stars in Cyg\,OB2 were originally concentrated
in a region of radius of $<1$ pc (that is consistent with the
observation that the initial radii of young clusters
are $\la 1$ pc; Kroupa \& Boily \cite{kr02}). It is therefore plausible
that at the moment of ejection of BD+$43\degr \, 3654$ the
stellar number density in the core of Cyg\,OB2 was high
enough to ensure that close encounters between its constituents were
frequent. The necessary condition for effective production of
runaways is a high binary fraction among massive stars. The recent
radial velocity survey of Cyg\,OB2 by Kiminki et al. (\cite{ki07}) gives a
lower limit on the massive binary fraction of $30-40 \%$,
while the comparison of the data from the survey with the
expectations of the Monte Carlo models suggest that this fraction
could be $\geq 80 \%$ (Kobulnicky \& Fryer \cite{ko07}). Thus we believe 
that $\sim 2$ Myr ago 
the conditions in the core of Cyg\,OB2 were favourable for the
dynamical processes discussed in Sect.\,4.

\section{BD+43$\degr$ 3654: binary-supernova scenario}

According to the binary-supernova scenario, a massive star residing
in a binary system could attain a high peculiar velocity due to the
disruption of the binary after the companion star exploded as a
supernova (Blaauw \cite{bl61}; Stone \cite{st91}). We show that the peculiar
velocity of BD+$43\degr \, 3654$ cannot be accounted for
within the framework of this scenario (cf. Vanbeveren et al. \cite{va07}).

The kinematic age of BD+$43\degr \, 3654$ of $\sim 1.8$ Myr and
the minimum possible lifetime of the supernova progenitor star of
$\sim$ 2.5-3 Myr imply that the actual age of BD+$43\degr \, 3654$ 
should be $\ga$ 4.5-5 Myr. The discrepancy between the
`observed' and the inferred ages could be reconciled if
BD+$43\degr \, 3654$ is a rejuvenated star. For example, it could
be rejuvenated through mass transfer from the primary star during
the Roche lobe overflow stage (e.g. Dray \& Tout \cite{dr07}). In this
case, the effect of rejuvenation would be significant only if the
mass gained by the rejuvenated star was larger than its initial
mass. Although one cannot exclude this possibility, we note that
stars more massive than $\sim$ 40 M$_{\odot}$ could lose a
significant fraction of their mass via the heavy stellar wind and
the Roche lobe overflow will not occur (Vanbeveren et al.
\cite{va98}). Another possibility is that BD+$43\degr \, 3654$ was
formed in the course of a close encounter between two
massive binaries, during which two stars of the binaries merged into
a single rejuvenated star (now BD+$43\degr \, 3654$) and caught a
third (more massive or more evolved) star to form a new binary,
while the fourth star was ejected as single (cf. Leonard \cite{le95}).

Let us assume that at the moment of supernova explosion in a binary
system the mass of the second star was 70 M$_{\odot}$. It is
obvious that to disrupt such a massive system, the explosion should 
be asymmetric so that the supernova stellar remnant attained a kick. 
In this case, the stellar remnant can
impart some momentum to the companion star in the course of
disintegration of the binary (Tauris \& Takens \cite{ta98}). The magnitude
of the momentum depends on the angle between the kick vector and the direction of motion of the exploding star and reaches its maximum for 
a certain value of the angle [given by Eq.\,(4) in Gvaramadze \cite{gv06}]. One can show that to accelerate a star as massive as BD+$43\degr \, 3654$ to the velocity of $40 \, {\rm km} \, {\rm s}^{-1}$, the kick direction should be very carefully tuned (i.e. should be within several degrees of
the direction towards the companion star), while
the magnitude of the kick should be very large ($\geq 700 \, {\rm km} \,
{\rm s}^{-1}$, if the stellar remnant is a neutron star,
or $\geq 200 \, {\rm km} \, {\rm s}^{-1}$, if the remnant is a black hole of mass $\sim$ 5 M$_{\odot}$). Although one cannot exclude this
possibility, we consider it as highly unlikely (cf. Gvaramadze 2006; 
Paper\,I).

\section{BD+43$\degr$ 3654: dynamical ejection scenario}

The numerical experiments by Leonard (\cite{le95}) showed that a
significant fraction of unbound blue stragglers (formed via
binary-binary encounters in dense clusters) attain peculiar
velocities large enough ($\geq 30 \, {\rm km} \, {\rm s}^{-1}$) to
be classified as runaways. It is therefore tempting to consider the
possibility that BD+$43\degr \, 3654$ is a blue straggler 
formed through an encounter between two tight massive binaries.
The most common outcome of encounters between such binaries is the
exchange of the more massive components into a new eccentric binary
or a single merged star and ejection of the less massive ones with
high velocities. The binary would ultimately coalesce into a single 
star if its orbit is sufficiently compact. 

Let us assume that BD+$43\degr \, 3654$ is the result of a merging
of two main-sequence (MS) stars of mass $M_1$ and $M_2$ ($M_1 \geq M_2$)
and of the same age $t_{\rm merg} \simeq 3$ Myr. Under these
assumptions, the merger product is also an MS star (of
mass M$_{\ast} \simeq$ M$_1$ + M$_2$), and its new age is given by (see
Meurs \& van den Heuvel \cite{me89}; Portegies Zwart et al. \cite{po99}):
\begin{equation}
t_{\ast} \sim {M_1 \over M_{\ast}} \, {t_{\rm MS} (M_{\ast}) \over
t_{\rm MS} (M_1)} \, t_{\rm merg} \, ,
\end{equation}
where $t_{\rm MS}$ (M$_{\ast})$ and $t_{\rm MS} (M_1)$ are the MS
lifetimes of stars of mass $M_{\ast}$ and $M_1$.
One can consider two cases: (i) BD+$43\degr \, 3654$ is the
product of a physical collision of two stars during the close
binary-binary encounter and (ii) BD+$43\degr \, 3654$ is the
result of coalescence of two stars in a close binary system. In the
first case, one should require that $t_{\ast} \sim 0$, since the
current (apparent) age of BD+$43\degr \, 3654$, $t=t_{\ast} +
t_{\rm kin}$, is comparable to its kinematic age $t_{\rm kin} \simeq 
1.8$ Myr. This requirement could be fulfilled
only if the encounter between binaries occurs very soon after their
birth in the association (i.e. $t_{\rm merg} \sim 0$ Myr), which
contradicts our assumption that $t_{\rm merg} \sim 3$ Myr. Note,
however, that here we neglected the possibility that the colliding
stars were already rejuvenated through mass transfer
in the original tight binaries so that their apparent age was much
less than the actual one. In the second case, one can assume that
the binary components coalesced into a single star only recently,
i.e. the time elapsed since the formation of BD+$43\degr \, 3654$
is less than the kinematic age of the binary. As an example,
suppose that the runaway binary consists of two stars of mass $\sim 35 \,
M_{\odot}$. For $t_{\rm MS}$ (35 M$_{\odot}) \simeq 4.5$ Myr and
$t_{\rm MS}$ (70 M$_{\odot}) \simeq 2.5$ Myr (Meynet \& Maeder \cite{me03}), 
and assuming that the
binary merged $\sim 1$ Myr after ejection from Cyg\,OB2 (i.e.
$t_{\rm merg} \sim 4$ Myr), one has from Eq.\,(1) that $t_{\ast}
\sim 1.1$ Myr. During the next $\sim 0.7$ Myr ($=t_{\rm kin}
-t_{\ast}$), the merged star evolves into a blue supergiant
with parameters similar to those of BD+$43\degr \, 3654$ [see the
evolutionary tracks by Meynet et al. (\cite{me94}) and the calibration of
parameters of Galactic O stars by Martins et al. (\cite{ma05})].

In Sect.\,1, we mentioned that the origin of two pulsars, B2020+28 and B2021+51, could be associated with Cyg\,OB2. In Paper\,I, we suggested that these pulsars are the remnants of runaway stars ejected (with velocities similar to those of the pulsars) from Cyg\,OB2 due to the dynamical three- or four-body encounters.
Our suggestion was based on the recently recognized
fact that the high-velocity pulsars could be the descendants of
high-velocity runaway stars (i.e. the peculiar velocities 
of pulsars do not necessarily originate from asymmetric 
supernova explosions; Paper\,I; Gvaramadze et al. \cite{gv08}). 
Strong support for this possibility comes from the 
discovery of early B-type stars moving with velocities of
$\sim 500-700 \, {\rm km} \, {\rm s}^{-1}$ (Edelmann et al. \cite{ed05};
Przybilla et al. \cite{pr08}; Heber et al. \cite{he08}). 
In Paper\,I, we used the similarity
between the spin-down and the kinematic ages of B2020+28 and 
B2021+51 to suggest that their progenitors were the
short-lived ($<$ 1 Myr) helium cores of massive stars, while from
the age of Cyg\,OB2 at the moment of ejection of the helium cores
($\sim$ 3 Myr) we inferred that the zero-age MS masses of the 
ejected stars were $\geq$ 50-60 M$_{\odot}$; it is believed 
that stars of this initial mass could leave behind a neutron star
(see Woosley et al. \cite{wo95}; Muno et al. \cite{mu06}; Bibby et al. \cite{bi08}).
The relative position of the pulsars and BD+$43\degr \, 3654$ on
the sky (see Fig.\,1) and the similarity between their kinematic
ages suggest that the three objects might have had a common origin in
the close encounter between two massive binaries.
\begin{figure}
\includegraphics[width=8cm]{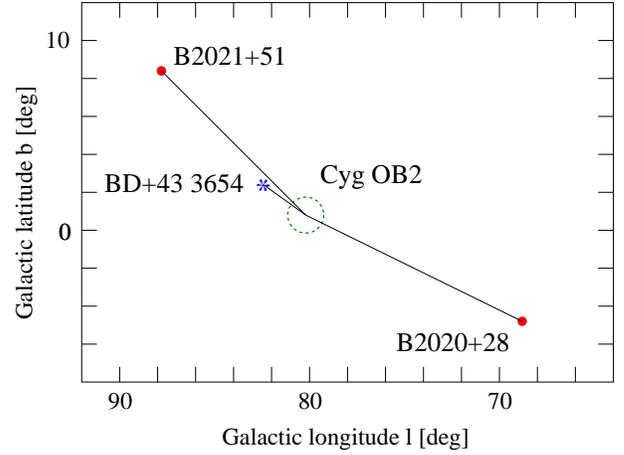}
\centering
\caption{Relative position of BD+$43\degr \, 3654$ and the pulsars
PSR~B2020+28 and PSR~B2021+51 on the sky. The circle of angular diameter
of $2\degr$ indicates the boundaries of the Cyg\,OB2 association.
}
\end{figure}
Below, we discuss this possibility in detail.

The (transverse) recoil velocity attained by the merged star is given by:
\begin{equation}
V_{{\ast},{\rm tr}} = M_{\ast} ^{-1} \left[ (m_1 v_{1,{\rm tr}})^2 +
(m_2 v_{2,{\rm tr}})^2 + 2m_1 m_2 v_{1,{\rm tr}} v_{2,{\rm tr}}
\cos\alpha\right]^{1/2} \, ,
\end{equation}
where $m_1, m_2$ and $v_{1,{\rm tr}}, v_{2,{\rm tr}}$ are, respectively,
the masses and the transverse velocities of the helium cores 
(the progenitors of the pulsars) and $\alpha$ is the angle between the ejection velocity vectors of the helium cores. We assume that the pulsars did not received (significant) kicks at birth and therefore move (almost) in the same direction as their progenitors. Adopting that the binary-binary encounter occured 1.8 Myr ago and neglecting the effect of
the Galactic gravitational potential, one obtains the transverse velocities of B2020+28 and B2021+51 as 
$v_{1,{\rm tr}} \simeq 200  \, {\rm km} \, {\rm s}^{-1}$ and 
$v_{2,{\rm tr}} \simeq 170 \, {\rm km} \, {\rm s}^{-1}$, and
$\alpha \simeq 160\degr$. Using Eq.\,(2), one
can show that to explain the `observed' transverse
velocity of BD+$43\degr \, 3654$ of $40 \, {\rm km} \, {\rm s}^{-1}$, the progenitor of B2021+51 should already lose most of its mass via the stellar wind, i.e. by the moment 
of binary-binary collision $m_2$ should be several times smaller than 
$m_1$. For $M_{\ast} =70\pm 15 M_{\odot}$ 
and assuming that $m_1 =(3-4)m_2$, one has from Eq.\,(2) that $m_1 \simeq (17-19)\pm 4 \, M_{\odot}$ (i.e. a quite reasonable figure). 
For the above parameters, one finds that the velocity vector of the
recoiled merged star is somewhat misaligned ($\sim 10\degr$) to the
residual velocity vector of BD+$43\degr \, 3654$. To explain this
misalignment, one can assume that Cyg\,OB2 has a peculiar (transverse) velocity of $\sim 7 \, {\rm km} \, {\rm s}^{-1}$ in the northwest direction (cf. Hoogerwerf et al. \cite{ho01}). 
The origin of peculiar velocity of this magnitude (typical of the OB associations near the Sun; de Zeeuw et al. \cite{de99}) could be understood if formation of Cyg\,OB2 was triggered by the collision between two molecular clouds (cf. Schneider et al. \cite{sc06}).

One can also constrain the radial velocity of BD+$43\degr \, 3654$
using the parallactic distances to B2020+28 and B2021+51,
respectively, of
$2.7_{-0.7} ^{+1.3}$ and $2.0_{-0.2} ^{+0.3}$ kpc (Vlemmings et al. \cite{vl04}).
Taken at face value, these distances imply the pulsar radial velocities
$v_{1,{\rm r}} \simeq 530_{-370} ^{+690} \, {\rm km} \, {\rm s}^{-1}$ and
$v_{2,{\rm r}} \simeq 160_{-110} ^{+160}  \, {\rm km} \, {\rm s}^{-1}$,
while from the conservation of the linear momentum 
one has $V_{\ast ,{\rm r}} \simeq -(150_{-110} ^{+200} ) \, 
{\rm km} \, {\rm s}^{-1}$. The peculiar radial velocity of Cyg\,OB2 of
$\simeq -6 \, {\rm km} \, {\rm s}^{-1}$ (derived from the mean systemic
velocity of the association of $\simeq -10 \, {\rm km} \, {\rm s}^{-1}$;
Kiminki et al. \cite{ki07}) only slightly changes the above figures. Our scenario
for the origin of BD+$43\degr \, 3654$ therefore suggests that the radial
component of the peculiar velocity of this star should be negative and at 
least as large as the transverse one. It could be that the radial velocity
measurements for BD+$43\degr \, 3654$ will invalidate our scenario, so that the origin of the runaway star and the pulsars would not be
related to each other. Even in this case, we believe that the origin of BD+$43\degr \, 3654$ should be accompanied by the ejection of two high-velocity stars, either early type B-stars or stripped helium cores 
of the more massive stars.

Thus, we suggest that the runaway massive star BD+$43\degr \, 3654$
originate from a close encounter between two binaries, originally 
consisting of a $\sim 35$ and a $\sim 50-60 \, M_{\odot}$ star. The more massive stars in each system evolved for $\sim 3$ Myr, losing most of their
mass, becoming helium stars. At this point, the binaries interacted, ejecting the helium stars at high velocity and resulting in the 
merger or near-merger of the $35 \, M_{\odot}$ stars into a $70 \,
M_{\odot}$ star, which recoiled at a proportionally lower velocity. The
more massive object is now seen as a blue straggler O4If star, while the
helium stars exploded as supernovae soon after the ejection and produced
the pulsars with a small or no kick at birth. 

\begin{acknowledgements}
We are grateful to the anonymous referee for constructive criticism and suggestions allowing us to improve the paper. VVG is grateful to A.\ Bogomazov, P.\ Kroupa and F.\ Martins for useful discussions and K.\ Weis for critically reading the manuscript. 
The authors acknowledge financial support from the 
Deutsche Forschungsgemeinschaft (grants 436 RUS 17/104/06 
and BO 1642/14-1) for research visits of VVG at the Astronomical 
Institute of the Ruhr-University Bochum.
\end{acknowledgements}

\end{document}